\documentclass[
journal=ancac3,manuscript=article]{achemso}

\usepackage[version=3]{mhchem} % Formula subscripts using \ce{}
\usepackage{graphicx}  % needed for figures
\usepackage{amsmath} %needed for math
\usepackage{dcolumn}   % needed for some tables
\usepackage{bm}        % for math
\usepackage{amssymb}   % for math
\usepackage{epstopdf}
\usepackage{xcolor} %for coloured text
\usepackage{hyperref}
\usepackage[normalem]{ulem}

\def \BSTS {BiSbTe$_{1.25}$Se$_{1.75}$~}
\def \Nbs {NbSe$_2$}

\author{Abhishek Banerjee$^1$, Ananthesh Sundaresh$^1$, R. Ganesan$^1$ and P. S. Anil Kumar$^1$}
\affiliation{\small{$^1$Department of Physics, Indian Institute of Science, 
Bengaluru 560012, India \\}}
\email{anil@iisc.ac.in}

\title{Signatures of Topological Superconductivity in Bulk Insulating Topological
Insulator \BSTS in Proximity with Superconducting \Nbs}

\begin{document}

\begin{abstract}
The combination of superconductivity and spin-momentum locking at the interface between an s-wave superconductor and a three-dimensional topological insulator (3D-TI) is predicted to generate exotic p-wave topological superconducting phases that can host Majorana fermions. However, large bulk conductivities of previously investigated 3D-TI samples and Fermi level mismatches between 3D bulk superconductors and 2D topological surface states have thwarted significant progress. Here we employ bulk insulating topological insulators in proximity with two-dimensional superconductor \Nbs~assembled {\it via} Van der Waals epitaxy. Experimentally measured differential conductance yields unusual features including a double-gap spectrum, an intrinsic asymmetry that vanishes with small in-plane magnetic fields and differential conductance ripples at biases significantly larger than the superconducting gap. We explain our results on the basis of proximity induced superconductivity of topological surface states, while also considering possibilities of topologically trivial superconductivity arising from Rashba-type surface states. Our work demonstrates the possibility of obtaining p-wave superconductors by proximity effects on bulk insulating TIs. 
\end{abstract}
\vspace{0.5cm}
\vspace{0.5cm}

Proximity effects between topological insulators and superconductors have attracted significant attention as potential sources of unconventional superconductivity. Right after the discovery of three dimensional topological insulators~\cite{TIreview1,TIreview2,TIreview3,TIreview4,TIreview5} it became evident that inducing superconductivity into two-dimensional surface states of 3D-TIs~\cite{Fu2008,Fu2009, TIreview2, TI_SC_JJ}  could lead to p-wave superconductivity~\cite{volovik1999fermion, senthil2000quasiparticle,read2000n,kitaev2001unpaired}, where defects in the form of edges or vortices host Majorana fermion states. In a different vein, it was also shown that inducing superconductivity into the `bulk' of topological insulators could lead to 3D topological superconductors whose surfaces could host surface Andreev bound states: essentially two dimensional analogues of linearly dispersing Majorana fermions~\cite{CuBi2Se3_1,CuBi2Se3_2,Odd_parity_TI_SC_1,Odd_parity_TI_SC_2,
Odd_parity_TI_SC_3}. Both these directions have been pursued vigorously with encouraging results including purported demonstrations of surface Andreev bound states in Cu$_x$Bi$_2$Se$_3$~\cite{CuBi2Se3_1,CuBi2Se3_2} and similar materials~\cite{sasaki2012odd}, proximity effects on 3D-TI/superconductor interfaces~\cite{proximity1, proximity2, proximity3, proximity4, proximity5, proximity6, Bi2Se3_NbSe2_1, TI_SC_1}, Majorana zero modes in vortex cores of Bi$_2$Te$_3$/NbSe$_2$~\cite{xu2015experimental, Majorana_TI} heterostructure, chiral 1D Majorana modes in Nb/quantum anomalous Hall insulator heterostructures~\cite{Majorana_TI_SC}. Yet, a lot of these results remain ambiguous and can also be interpreted as consequences of more trivial effects. Experimental platforms that can manifest clear indications of topological superconductivity or Majorana fermions in TIs are therefore highly sought after.

One of the primary sources of such ambiguity is the large bulk conductivity in first generation topological insulators like Bi$_2$Se$_3$. Most experiments on topological proximity effects have been performed on bulk conducting TIs~\cite{proximity1, proximity2, proximity3, proximity4, proximity5, proximity6, Bi2Se3_NbSe2_1, TI_SC_1,xu2015experimental, Majorana_TI}, and it is not clear whether the reported effects have topological origin. To remove these ambiguities it is imperative to induce superconductivity into topological surface states of {\it bulk insulating} TIs, for instance of the Bi$_{x}$Sb$_{2-x}$Te$_{y}$Se$_{3-y}$ class. Such TIs achieve almost fully surface state dominated conduction and have sparked off a panoply of recent breakthroughs in condensed matter physics~\cite{QAHE1,axion_insulator,Majorana_TI,QHTI2}. However, proximity effect between a superconductor and a bulk insulating topological insulator remains difficult to achieve. Reduced bulk carrier densities and large Fermi surface mismatches between bulk three dimensional superconductors like Nb and  intrinsically two-dimensional topological surface states leads to weakly induced superconducting correlations. The inherent lack of surface bonding sites because of the layered nature of these materials may prevent strong chemical bonding with the superconductor, further exacerbating the problem. This may explain recent results on Nb/BSTS interfaces where electron-electron correlations rather than superconducting correlations appeared to dominate electrical transport~\cite{stehno2017conduction,tikhonov2016andreev}. 

Several things must therefore conspire for topological superconductivity to occur in bulk insulating TIs: i) The host TI should be able to support large surface carrier density whilst maintaining negligible bulk carrier density ii) The chemical identities of the TI and SC must be similar to create a homogeneous interface that allows strong wave-function overlap on either side iii) The Fermi surfaces on the SC and TI side should be well-matched to allow momentum conserved tunneling of quasiparticles. To this end, we fabricate van der Waals junctions between {\it bulk-insulating} topological insulator \BSTS (BSTS) and two-dimensional s-wave superconductor NbSe$_2$ and study their electrical transport properties. The chemical similarity between NbSe$_2$ and \BSTS, the intrinsic two-dimensional Fermi surfaces of both materials and highly bulk insulating nature of \BSTS all work in our favour. We unveil several signatures that are consistent with proximity induced superconductivity of the topological surface states including a differential conductance spectrum featuring two superconducting gaps, an intrinsic conductance asymmetry that vanishes when the induced superconductivity is quenched by an in-plane magnetic field and perhaps most strikingly, ripple-like features in the differential conductance spectra at biases considerably larger than the superconducting gap that vanish when bulk superconductivity is killed.
   
\begin{figure}[!t]
\includegraphics[width=1.\linewidth]{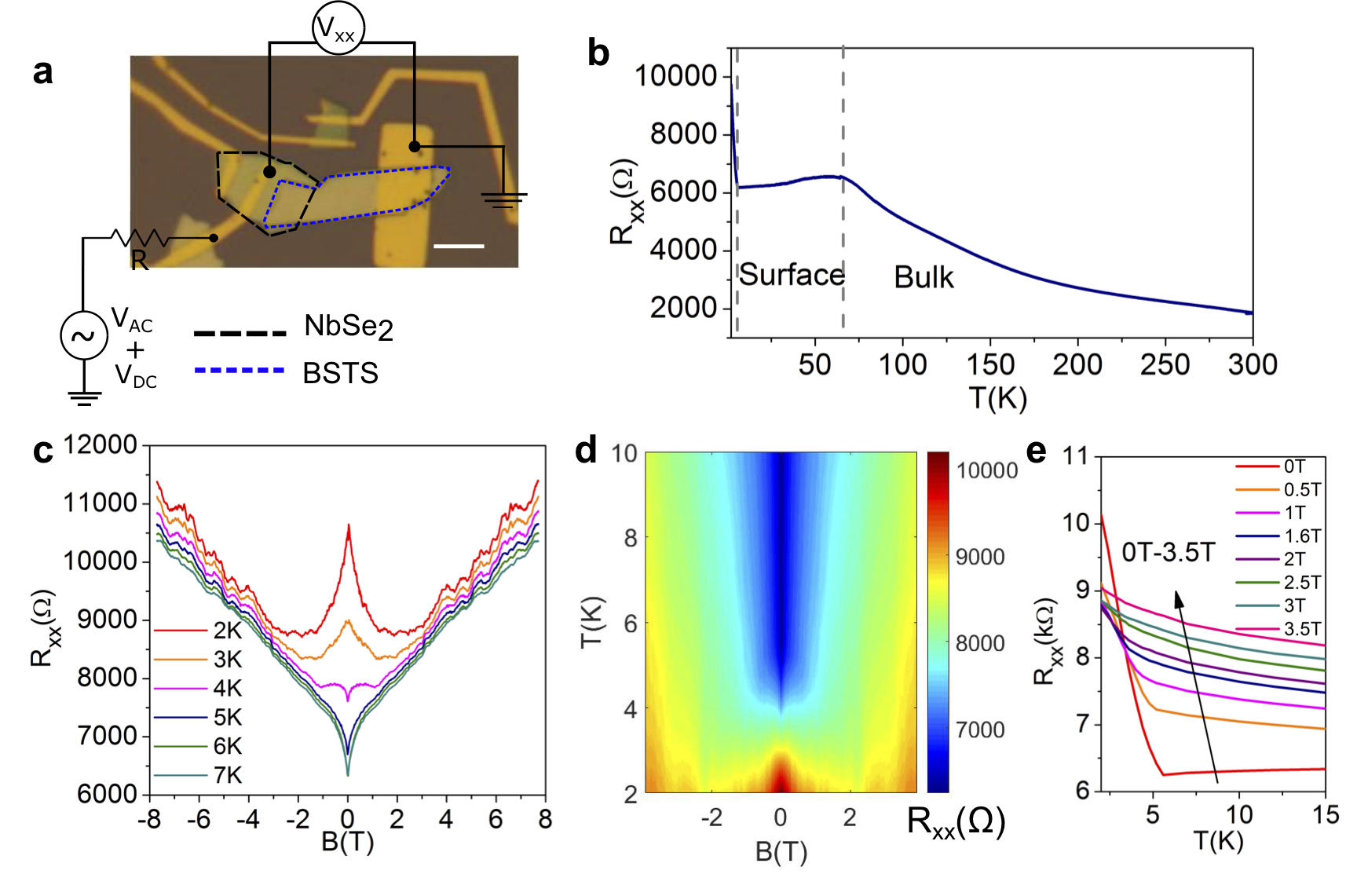}
\caption{(a) Optical micrograph of a \Nbs/\BSTS junction. Scale bar 5 $\mu$-m. (b) Resistance {\it vs} temperature showing surface dominated conduction (c) Magnetoresistance measurements at different sample temperatures (d) 2D color plot showing resistance as a function of magnetic field and temperature across the superconducting transition (d) R-{\it vs}-T showing the superconducting transition at different magnetic fields}
\label{fig01}
\end{figure} 

\section{Results and Discussion}

{\bf Device fabrication and zero bias electrical transport}

 \BSTS samples are obtained from the same crystal that has been used in our previous works to demonstrate large surface state transport, vanishing bulk conductivity and quantum Hall effects~\cite{Banerjee16,banerjee2018quantized}. \Nbs~flakes are obtained from \Nbs~single crystals with bulk $T_c=7.1K$. Heterostructures of  NbSe$_2$/BSTS are prepared using an all-dry van der Waals transfer method. Subsequently, e-beam lithography followed by e-beam evaporation of Cr/Au:10nm/100nm is used to define electrical contacts to the sample. Details of device fabrication and single crystal characterization can be found in supplementary material section A and B. Electrical transport experiments are performed using a low-frequency AC lock in method. 10nA-20nA of AC bias currents are used throughout the experiments. For differential conductance measurements, the DC bias is generated using a 16-bit digital-to-analog converter, and is added to the AC bias within a home-built setup. All measurements reported here use a current biased scheme.

~\ref{fig01}(a) depicts the optical micrograph of a typical device. Of the total 5 electrical contact lines, two contact the \Nbs~layer and one large contact pad connects with the \BSTS layer. An extra pair of contact lines connect an adjoining \Nbs~layer that is electrically isolated from the hetero-junction device, but can be used to calibrate the device temperature by measuring its superconducting transition. A large area of overlap is maintained between the BSTS and \Nbs~layers. Resistance {\it versus} temperature(R-T) measurement (~\ref{fig01}(b)) at zero magnetic field shows characteristics of a typical bulk insulating topological insulator. For T$>80K$, R-T shows an insulating behavior originating from the gapped bulk states. The low temperature range ($5.6K<T<80K$) shows metallic transport arising from the topological surface states. For T$<5.6K$ we observe a sharp non-saturating increase in sample resistance, indicating an onset of superconductivity. The {\it rise} in resistance indicates that our samples are tuned into the so-called tunneling regime where Andreev reflection is suppressed due to the high bulk resistivity of our samples. This is essential to our experiments, since the tunneling regime allows an accurate measurement of the density of states at the \Nbs/TI interface.

Magnetoresistance measurements, depicted in ~\ref{fig01}(c), show a large zero-field resistance peak at T=2K. With increase in magnetic field, the resistance peak is suppressed showing negative magnetoresistance. This suppression continues till B$\sim$3.5T after which positive magnetoresistance arising from weak-anti-localization(WAL) of topological surface states takes over. With increase in sample temperature, the negative MR peak gets suppressed, and a positive MR dip due to WAL becomes dominant and persists till very large temperatures(T $\simeq$ 30K). The suppression of the zero-field peak is clearly visible in the 2D resistance map shown in ~\ref{fig01}(d). ~\ref{fig01}(e) shows a sequence of R-T measurements at different magnetic fields. The zero field measurement depicts a strong resistance upturn for T$\leq5.6K$. Note that this temperature is lower than the transition temperature of bulk \Nbs($T_c=7.1K$). At larger magnetic fields, the resistance upturn is suppressed while the kink in R-T curves(at $T=5.6K$ at $B=0T$) shifts towards lower temperatures, and becomes invisible at B$\simeq3T$. The field scale of B$=3T$ roughly corresponds with the upper critical field H$_{c2}=3.5$T of NbSe$_2$.
 
\begin{figure}[!t]
\includegraphics[width=1.\linewidth]{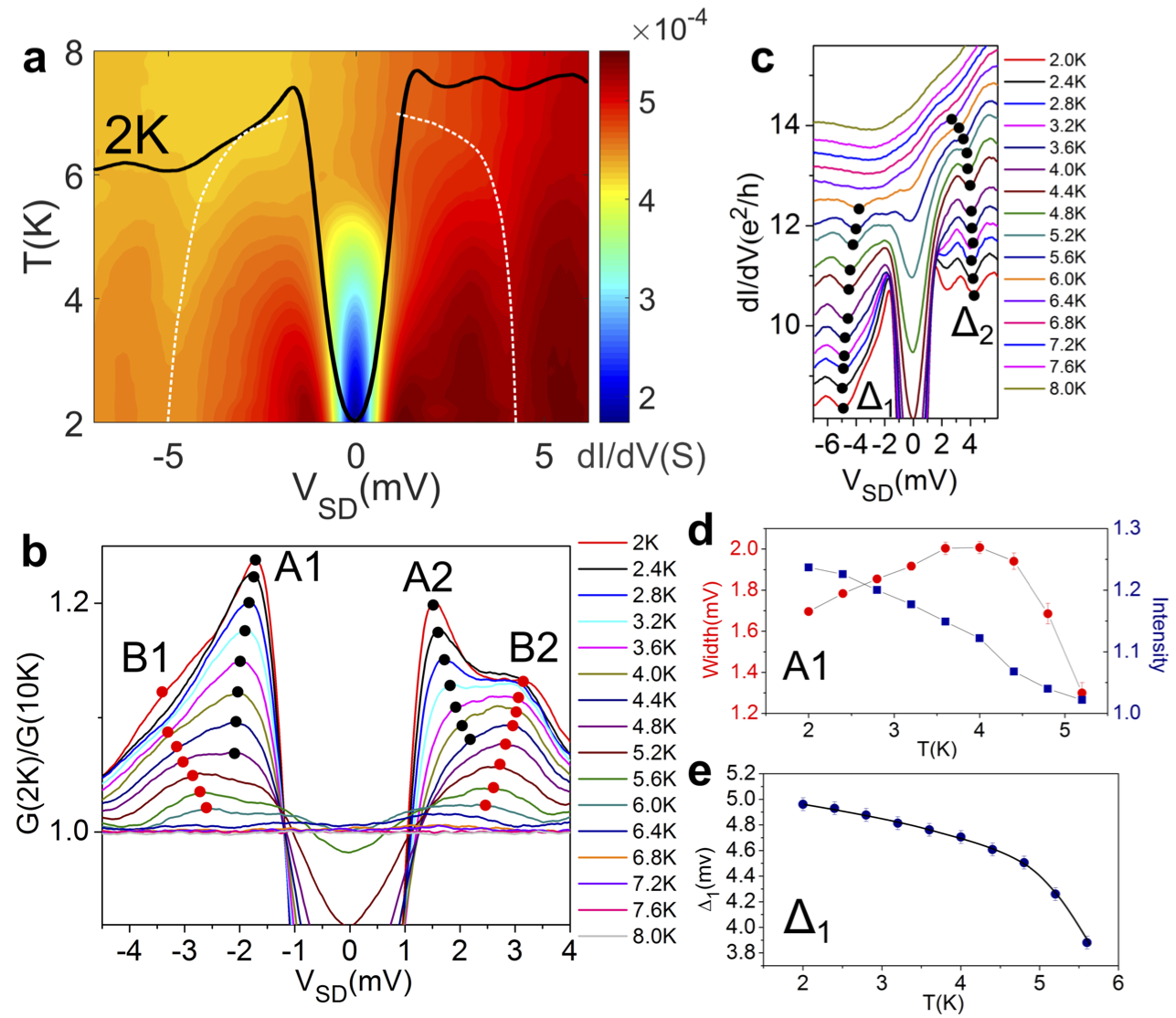}
\caption{(a) 2D color plot of differential conductance as a function of temperature and source-drain DC voltage bias $V_{SD}$ (b) Normalized differential conductance G(2K)/G(10K) showing a two-gap structure. The black and red dots track the evolution of coherence-like peak A1(A2) and B1(B2) respectively (c) Differential conductance ripples $\Delta_1$ and $\Delta_2$ that exist at biases larger than the superconducting gap. Black dots trace their evolution with increasing temperature. (d) Evolution of the width and intensity of peak A1 with increasing temperature (e) Evolution of position of the conductance dip at $\Delta_1$ with increasing temperature.}
\label{fig02}
\end{figure} 

{\bf Differential conductance measurements}

%\subsection{Differential conductance measurements}
To probe the presence of proximity effects, we perform two-terminal differential conductance measurements employing a two-probe measurement scheme in a current biased configuration as depicted in ~\ref{fig01}(a). The DC current biased differential conductance data are converted to the DC voltage biased differential conductance data, by taking into account the resistance of the un-proximitized BSTS layer {\it not} contacted by \Nbs, that appears in series with the BSTS/NbSe$_2$ junction resistance. The large resistivity of our bulk insulating samples (3-7 k$\Omega$/sq) makes this lead resistance an important parameter that must be accurately estimated. Note that using a four-probe geometry cannot alleviate this problem since such a configuration would only remove the Au-BSTS and Au-NbSe$_2$ contact resistances ($\sim$200-500$\Omega$) which are much smaller than the lead resistance provided by the un-proximitized BSTS layer. We therefore resort to Blonder-Tinkham-Klapwijk(BTK) theory~\cite{BTK1} using the total lead resistance as a parameter to fit our experimental differential conductance data. We relegate the details of BTK fitting to supplementary material section C. This produces a lead resistance of $R_l \simeq 4 k\Omega$ and the superconducting gap $2\Delta V_{A} \simeq 2.4mV$, which agrees well with the superconducting gap of bulk \Nbs, $2\Delta_0 \simeq 2.2 mV$. The DC voltage bias differential conductance data are obtained from the DC current biased differential conductance data (provided in supplementary material section I) by subtracting the voltage drop across the contact resistance $R_l$, and also removing the contribution of $R_l$ to the measured differential conductance. We note that an error in estimation of the lead resistance R$_l$ could in-principle lead to an apparent shift in both the bias voltage and amplitude of various differential conductance features, resulting in possibly erroneous conclusions. However, our BTK fitting procedure determines the amplitude and temperature dependence of the superconducting gap $\Delta_0$ of bulk \Nbs with a high degree of accuracy (see supplementary material Figure S3(k)), indicating reasonable internal self-consistency. Furthermore, the saturation of \BSTS resistivity at low temperatures ($T<20$K) as observed clearly in the non-superconducting region of the R-{\it vs}-T data depicted in ~\ref{fig01}(b) and ~\ref{fig01}(e), and in our previous works~\cite{Banerjee16}, indicates that the lead resistance does not vary significantly in the temperature range of our interest (2K-7K), thereby minimizing the possibility of error in the estimation of $R_l$. Nonetheless, since the estimation of $R_l$ is a subtle issue, to aid the reader, we have provided the raw differential conductance data as a function of the DC voltage bias, without correcting for $R_l$ in supplementary material section I.

The differential conductance data provides striking signatures of superconducting proximity effect as depicted in the 2D color plot of ~\ref{fig02}(a) where a strong zero bias conductance dip and series of peak/dip features at finite biases are observed. To rule out non-superconducting features,we follow the standard practice of normalizing the differential conductance data with the spectra obtained at  $T=10K$ just above the superconducting transition temperature of \Nbs, as depicted in ~\ref{fig02}(b). We draw attention to several important features in the differential conductance spectra, before discussing each of these aspects in detail. 

i) First, the strong differential conductance dip at zero bias flanked by two sets of peaks (A1, A2 and B1, B2) indicative of two superconducting gaps. Such double gap features have been observed previously in several proximity effect studies~\cite{proximity5,proximity6}. While the larger gap (B1 and B2) corresponds to electron tunneling into the unproximitised part of the superconductor away from the TI/SC interface, the smaller gap is usually attributed to the inverse-proximity effect wherein the region of the SC in contact with the TI has reduced superconducting correlations. Equivalently, this gap can also be associated with electron tunneling into the topological surface state which has now acquired a superconducting gap {\it via} the proximity effect. With increasing sample temperature, the peak positions of A1(A2) and B1(B2) show opposite behavior. While the gap at B1(B2) decreases, the gap at A1(A2) expands  as plotted in ~\ref{fig02}(d), while the intensities of both peaks reduce. In fact, for $T >\simeq 4.8K$, A1(A2) completely vanishes and appears to merge with B1(B2). This further supports the proximity-induced origin of A1, in that it disappears well before B1 that corresponds to the parent superconductor. 

ii) Second, the dI-dV measurements show an intrinsic conductance asymmetry with respect to the sign of the DC bias voltage. The normalized conductance peak at A1 is considerably larger than its positive bias counterpart at A2. However, this asymmetry is less evident for peaks B1 and B2. In fact, the asymmetry disappears with increasing temperature and is roughly co-incident with the merger of peaks A1(A2) and B1(B2). For $T\geq4.4K$, the asymmetry is completely absent.

iii) Third, we observe several differential conductance ripples comprising consecutive dips and peaks at biases larger than the gap voltages. First, to ascertain whether these features have a superconducting origin or not, we analyze the temperature dependencies of the position of a pair of these dips with the strongest signature: $\Delta_1$ and its positive bias counterpart $\Delta_2$ appearing at $V_{SD}\simeq \pm 4-5$meV. These appear as the striking dip-like features marked as white dashed lines in ~\ref{fig02}(a) and shown as $\Delta_1$ and $\Delta_2$ in ~\ref{fig02}(c). With increasing temperature, the dip feature collapses (~\ref{fig02}(e)) and completely disappears at $T\simeq 6K$ indicating its superconducting origin. Such ripples have been previously reported in superconductors showing multi-phonon effects like Pb~\cite{multiphonon} and correlated superconductors with magnonic exictations~\cite{magnonsuperconductivity}. However, since the parent superconductor \Nbs~shows no such effects, we can rule out such explanations since superconductivity in our samples is proximity induced. We point out that similar super-gap features were recently observed in junctions of \Nbs~and bulk conducting Bi$_2$Se$_3$~\cite{Bi2Se3_NbSe2_1}. However, the authors did not focus on this effect and provided no explanations for its origin.   

{\bf Differential conductance under in-plane magnetic fields}

\begin{figure}[!t]
\includegraphics[width=1.\linewidth]{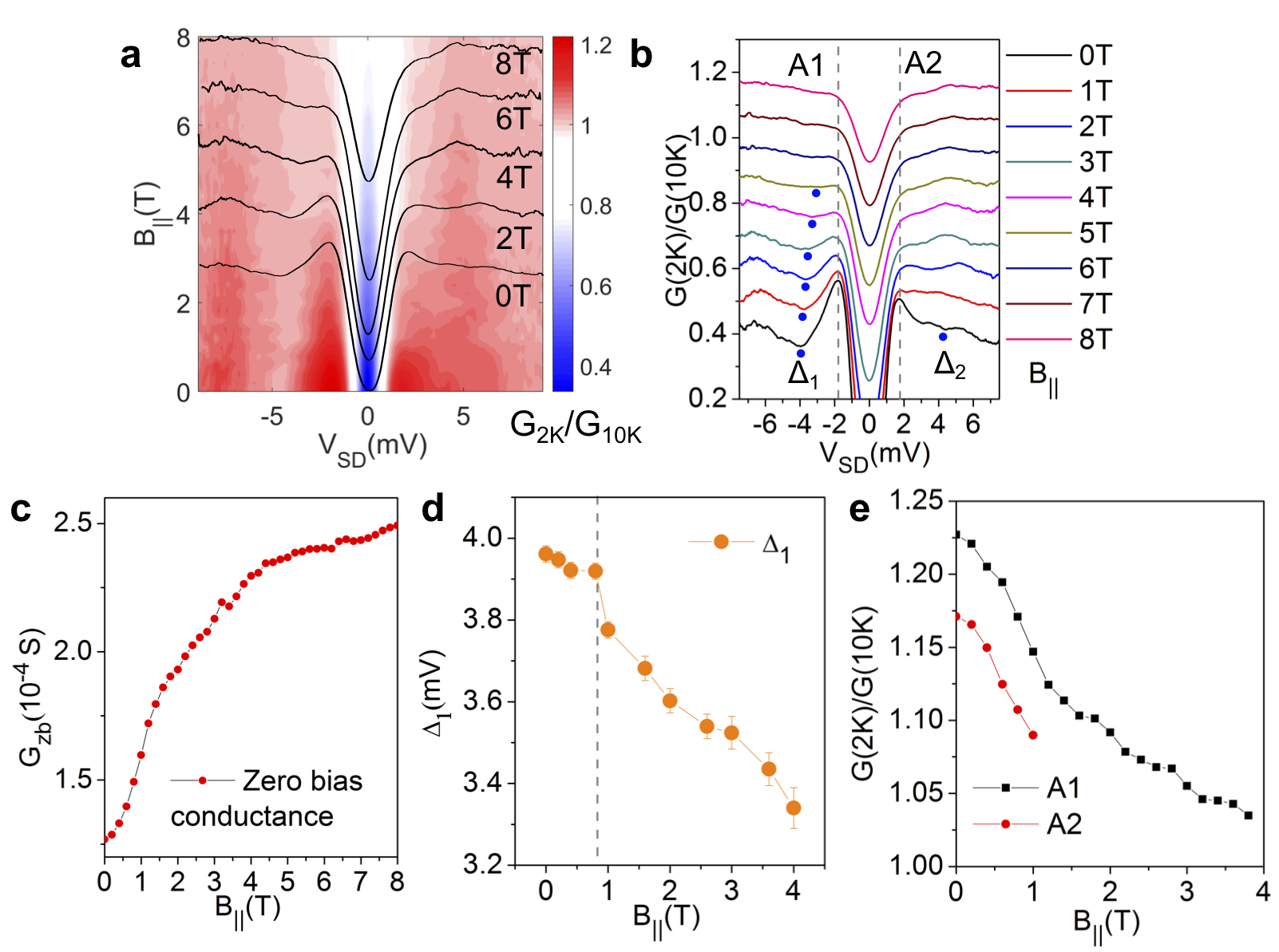}
\caption{(a) 2D color plot of normalized differential conductance G(2K)/G(10K) {\it vs} $V_{SD}$ as a function of in-plane magnetic field. (b) Evolution of G(2K)/G(10K) showing the breakdown of asymmetry with increasing in-plane magnetic field. The super-gap ripples at $\Delta_1$ and $\Delta_2$ marked as blue dots also disappear. (c) Zero bias conductance $G_{zb}$ as a function of in-plane magnetic field. (d) Evolution of bias position of dip at $\Delta_1$ with $B_{||}$. (e) Evolution of normalized coherence peak intensity at A1 and A2 with $B_{||}$.}
\label{fig03}
\end{figure}

We now explore how these features evolve upon the application of magnetic fields oriented parallel to the plane of the sample. As depicted in ~\ref{fig03}(a), the superconducting gap persists upto 8T, the largest magnetic field attainable in our system. However, an examination of the normalized differential conductance $G_{2K}/G_{10K}$ reveals that asymmetry between peaks A1 and A2 that is observed at low magnetic fields is completely extinguished for $B_{||} >\simeq 5T$. In fact, notice how the entire dI-dV spectra that is highly asymmetric at $B_{||}=0T$ becomes highly symmetric at $B_{||}=8T$. Additionally, the super-gap ripples, indicated as $\Delta_1$ and $\Delta_2$ in ~\ref{fig03}(b)), also vanish with in-plane magnetic field and are completely suppressed at 5T as shown in ~\ref{fig03}(d) for $\Delta_1$.  A critical value of $B_{||} \simeq 5T$ also appears in the evolution of the zero bias conductance($G_{zb}$) plotted in ~\ref{fig03}(c). With increasing magnetic field, $G_{zb}$ rises sharply until around 5T, and then almost saturates.

 We interpret this evolution as arising from two different superconducting orders: the low field differential conductance corresponds to a combination of tunneling into the proximity induced topological surface state and the bulk superconductor \Nbs. In general, increasing in-plane magnetic fields causes depairing of Cooper pairs due to orbital motion, thereby suppressing superconductivity. However, this effect is presumably stronger for the proximitized topological surface states, compared to two-dimensional \Nbs~where it is known that Ising pairing between Cooper pairs causes superconductivity to survive at in-plane fields much larger than that dictated by the Pauli susceptibility limit~\cite{Ising1}. Therefore, while proximity induced superconductivity and its consequent features including differential conductance asymmetry and super-gap ripples are suppressed completely by $B_{||}=5T$, the superconductivity in the Ising paired \Nbs~survives and produces a symmetric differential conductance spectrum as expected for an s-wave superconductor.   

The asymmetry in dI-dV therefore appears to be associated with the proximity induced superconducting order, rather than superconductivity of bulk \Nbs. The asymmetric spectrum has a natural explanation as a consequence of the linear density of states of the Dirac spectrum of topological surface states $D(E) \propto E$. When a superconducting gap is induced into the two-dimensional TSS, it retains its linear shape since the in-gap spectra must continuously match with the normal state spectra outside the gap. This is in contrast to bulk derived 2D Rashba type surface states that have finite mass and feature a constant density of states. The coherence peaks(A1 and A2) arising from the proximitisation of the TSS therefore retain an intrinsic asymmetry that vanishes when the sample temperature or magnetic field becomes large enough to close the TSS gap. Beyond such a temperature or magnetic field, only symmetric features originating from \Nbs~are visible.

{\bf Differential conductance under perpendicular magnetic fields}

\begin{figure}[!t]
\includegraphics[width=1.\linewidth]{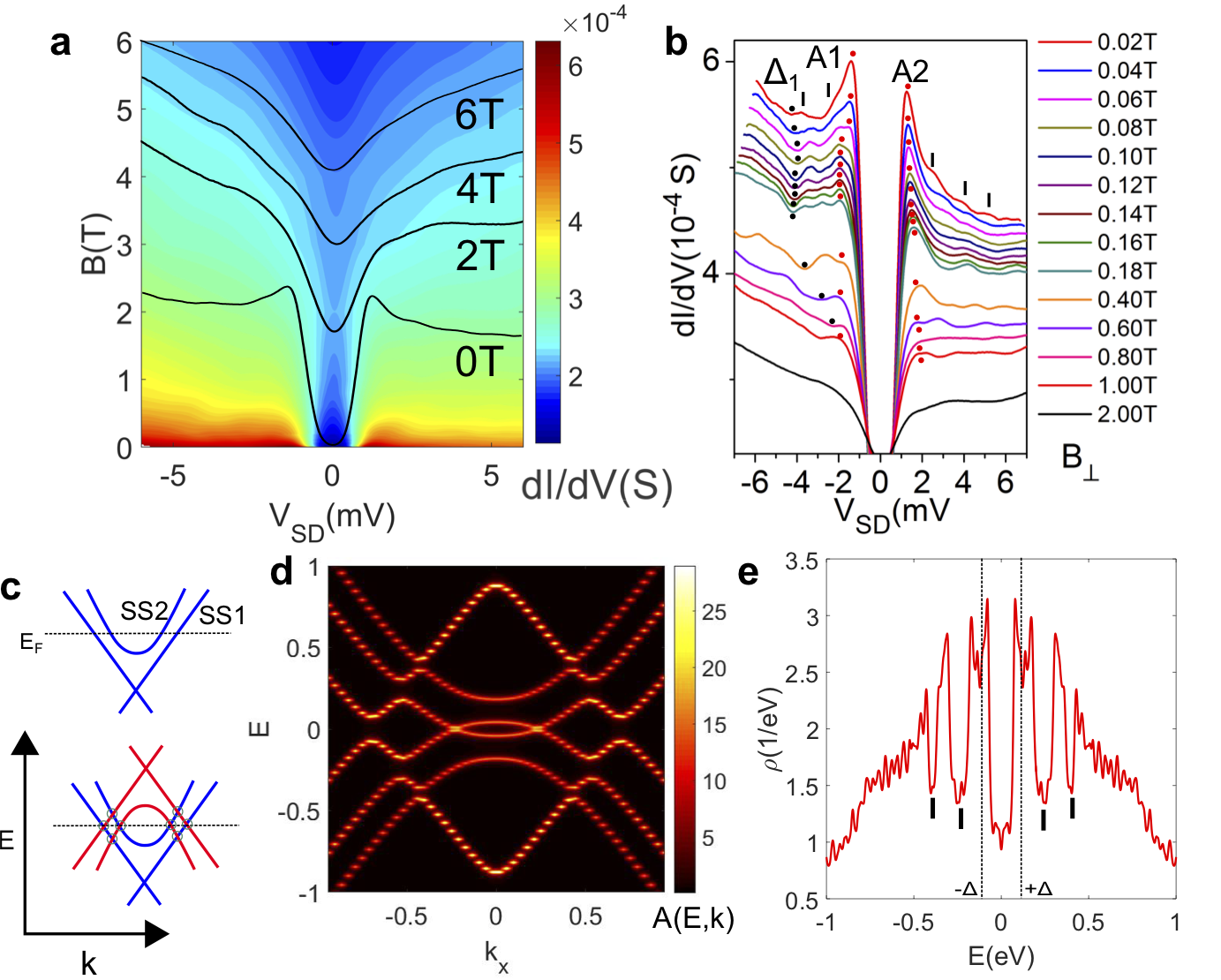}
\caption{(a) 2D color plot of differential conductance as a function of out-of-plane magnetic field from 0T to 8T. (b) Low field(0T-2T) evolution of differential conductance spectrum showing several ripples at super-gap biases marked by vertical solid lines  (c) (Upper panel) E-k band diagram of two surface confined states(SS1 and SS2) closely spaced in energy (Lower panel) Electron-hole excitation spectrum(no superconducting gap) showing band-crossings at the Fermi level and also at higher energies, indicative of interband Cooper pairing. (d) Spectral function evaluated for a multi-band topological superconductivity in a thin slab of a TI (e) Corresponding density of states showing mini-gaps at biases larger than the superconducting gap. }
\label{fig04}
\end{figure}

Now we investigate differential conductance measurements in the presence of perpendicular magnetic fields as shown in ~\ref{fig04}(a). We observe that superconductivity is suppressed at $B\simeq 3.5T$, detected as a change in the variation of the zero bias conductance dip with magnetic field as seen in ~\ref{fig04}(a). (see also ~\ref{fig01}(c)). For $0<B<\simeq 3.5T$, zero bias conductance is enhanced, indicating destruction of superconducting order. Beyond this field, zero bias conductance gets suppressed as the sample enters into a regime where transport is dominated by a competition between  a negative correction to conductance arising from electron-electron(e-e) interactions and a positive correction arising from weak-antilocalization (WAL). The WAL effect is verified independently using magnetoresistance measurements presented in ~\ref{fig01}(c) and (d). In topological insulators, these two corrections almost cancel each other out in zero magnetic field. In a finite magnetic field, the positive correction due to WAL is suppressed due to broken time-reversal symmetry, thereby unmasking the negative conductance correction from e-e interactions. The applied voltage bias acts like an effective temperature that weakens the negative correction from e-e interactions, thereby leading to positive differential-conductance as seen in ~\ref{fig04}(a) for $B>3.5T$. With increasing magnetic field, the zero bias conductance decreases as corrections from e-e interactions become more and more dominant. This distinction between superconductivity induced and e-e interaction induced differential conductance spectra is notable, particularly because previous experiments on superconducting proximity effects on bulk insulating TIs have been mired by strong differential conductance signals arising from e-e interactions rather than superconducting correlations~\cite{stehno2017conduction, tikhonov2016andreev}.

We now discuss the low-field measurements presented in ~\ref{fig04}(b). Specifically, we notice several differential conductance ripples above the superconducting gap that persist and even appear to get enhanced with small magnetic fields $B<0.2T$. For larger fields, the ripples become non-uniform and completely disappear before reaching $B \simeq 2T$. As we discussed before, these ripples cannot be explained as a consequence of multi-photon effects or strong correlations, none of which have been observed in extensive studies of \Nbs. Extensive experiments on the BSTS material class have also never revealed any signatures of magnetic ordering or correlations. A more reasonable explanation could be offered by considering the effects of low critical currents, as observed previously in point-contact spectroscopic measurements~\cite{Sheet}. It is possible that superconductivity induced into the topological surface states is rather weak, with critical currents as small as $I_c \simeq 1-3 \mu A$. Such a scenario would lead to a differential conductance dip when the current through the device exceeded the critical current, likely at a bias larger than the superconducting gap. However, this cannot explain the observation of multiple dips in our data. Additionally, such small critical currents are not consistent with disproportionately large magnetic fields($B \simeq 2T$) until which the ripples survive.

%{\bf Discussion}

An interesting explanation for this effect is obtained by considering multi-band Cooper pairing between different states residing on the TI surface. As shown schematically in ~\ref{fig04}(c), the topological surface states of a TI are usually accompanied by several closely spaced bands. Such bands may arise due to defect induced resonances~\cite{xu2017disorder}, bulk 2D sub-bands co-existing with the surface states at the chemical potential or band-bending induced spatial confinement of bulk states~\cite{WALAdditionalChannel,Banerjee16}. In the presence of induced superconducting order, multiple gaps may open up at finite energies away from the superconducting gap due to electron-hole mixing between the topological surface state and additional surface bands . To illustrate this effect we first evaluate the spectrum of Bogoliubov quasiparticles in a thin TI slab consisting of the topological surface state and a finite number of two-dimensional bulk bands, with the four possible types of superconducting order $\Delta_{1,2,3,4}$ allowed by symmetry (see supplementary material section E). 

To aid clarity we first choose theoretically convenient parameters, before describing a realistic simulation of a BSTS/NbSe$_2$ junction. In ~\ref{fig04}(d), we depict the spectral function $A(E,k)$ for an odd-parity $\Delta_2$ type superconducting order, showing the Majorana fermion {\it butterfly} spectrum at zero energy~\cite{Odd_parity_TI_SC_2,Odd_parity_TI_SC_4}. These in-gap states are the so-called surface Andreev bound states and essentially represent two-dimensional analogs of Majorana fermions~\cite{Odd_parity_TI_SC_1,Odd_parity_TI_SC_2,Odd_parity_TI_SC_3,Odd_parity_TI_SC_4,
TISC_review}. Accompanying this, we observe several finite-energy gaps corresponding to crossings of the topological surface states with the 2D bulk bands. The corresponding density of states is plotted in ~\ref{fig04}(e) and shows several dips (marked by solid vertical lines) beyond the superconducting gap ($\Delta \simeq 0.1$eV). Most spectacularly, it can be shown that the finite bias gaps appear only for those superconducting order symmetries that also lead to a Majorana fermion mode at zero energy. This can be understood intuitively as a redistribution of spectral weight: level repulsion at higher energies pushes states into the gap, leading to the  zero energy Majorana modes. Therefore, Majorana zero modes in such systems are always accompanied by mini-gaps at biases much larger than the primary superconducting gap. Additionally, the superconducting origin of the peak-dip features associated with these finite energy ripples may be gauged by evaluating their dependence on sample temperature. With increasing sample temperature, not only is the depth of the peak-dip structure reduced due to a reduction of the mini-gap amplitude, but also the voltage bias positions associated with these features are drawn closer to zero energy. This happens because the 2D-sub band levels which were originally pushed away to larger energies by level repulsion due to the primary superconducting gap, are now drawn closer towards zero energy as the primary gap collapses. This is demonstrated by detailed simulations of the peak-dip structure of the ripples with decreasing values of superconducting gap amplitude provided in supplementary material section F. This effect is exactly what is observed experimentally in ~\ref{fig02}(c) and ~\ref{fig02}(e) where the dip feature $\Delta_1$ moves towards zero energy by $\simeq 1$meV, which matches very well with the primary superconducting gap amplitude $\Delta_0 \simeq 1.1$meV due to \Nbs.

Further,  as shown in ~\ref{fig04}(e), at zero energy the density of states shows a large superconducting gap, and a small {\it dip} exactly at zero bias. This is despite the presence of a 2D Majorana mode and arises from the vanishing density of states of the linear spectrum of Majorana modes at zero energy~\cite{TISC_review}. This is of course contrary to the general belief that a Majorana mode must always lead to a conductance {\it peak} at zero bias. This is true only in one dimension; in higher dimensions the zero mode generically leads to a conductance dip and is therefore difficult to separate from conductance dips arising from non-topological effects~\cite{TISC_review,Odd_parity_TI_SC_4}. In such a scenario, the detection of superconducting conductance dips at large biases may offer an alternate test for topological superconductivity. 

\begin{figure}[!t]
\includegraphics[width=1\linewidth]{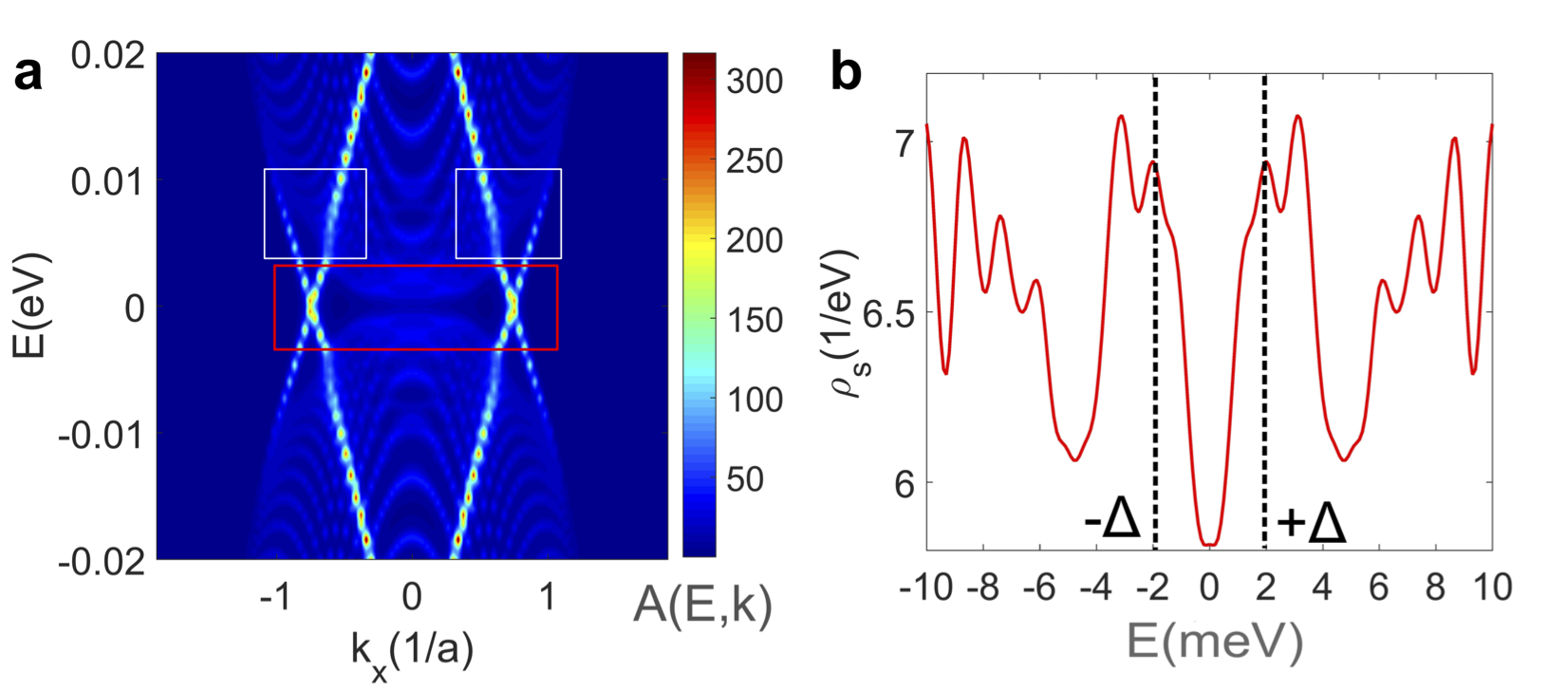}
\caption{(a) Surface spectral function evaluated for a realistic interface between NbSe$_2$ and a 50~QL thick slab of BSTS. The red box depicts the 2D Majorana type surface Andreev bound state. The white boxes depict the avoided crossings between the topological surface state and bulk derived 2D sub-bands (e) Corresponding density of states showing oscillations arising as a consequence of topological superconductivity. }
\label{fig05}
\end{figure} 

We now perform a realistic simulation of a hetero-junction between NbSe$_2$ and BSTS and evaluate the spectrum of surface superconducting states.  The work function of NbSe$_2$ is $\phi_{SC}\simeq$5.6-5.9eV~\cite{Liue1600069,shimada1994work} whereas that for BSTS is $\phi_{TI}\simeq$5.0-5.1eV~\cite{takane2016work}. The large work-function mismatch $\phi_{SC}-\phi_{TI}\simeq$0.5-0.9eV between the two materials leads to p-type doping of BSTS and forces the chemical potential of the BSTS layer to get pinned at the valence band maxima. To perform a realistic band-structure calculation for BSTS, we tune the model parameters to fit the experimentally measured angle resolved photoemission spectrum of BSTS obtained from our previous work (see supplementary material section H). In ~\ref{fig05}(a), we depict the surface spectral function of a $N_z=$50~QL thick BSTS slab with an induced superconducting gap $\Delta=2.0meV$. At zero energy, the {\it butterfly} spectrum arising from 2D surface Majorana fermions is clearly observed (red box in ~\ref{fig05}(a)). Simultaneously, the avoided crossing between the topological surface state band and the bulk derived 2D sub-bands (marked using white boxes in ~\ref{fig05}(a)) give rise to oscillations in the corresponding density of states as shown in ~\ref{fig05}(b) at biases that are significantly larger than the induced superconducting gap. The ripples observed in our calculations are separated by $\simeq$2-3meV in energy, exactly as observed in our experimentally measured differential conductance spectrum. Similar calculations are performed for BSTS slabs with thicknesses of $N_z=$20~QL and $N_z=$35~QL to obtain qualitatively similar results, but with larger energy scales separating the super-gap dI-dV ripples (see supplementary material section H). Qualitatively, the energy separation of the ripples can be estimated from the energy spacing between bulk-derived 2D sub-bands. In our experiments, the bulk bands originate from $J=1/2$ Se(Te) p-type atomic orbitals, which form the valence band of our system with an experimentally measured band-width of $\Delta_{VB} \simeq 200$ meV~\cite{lohani2017band} (see supplementary material section G). This results in a sub-band spacing $\Delta_{SB}=\Delta_{VB}/N_z \simeq 4$meV for a 50~QL layer slab. This simple estimation is in close agreement with the energy scale of 2-3meV derived from the Green's function based density of states calculations. The appearance of differential conductance ripples at large biases in our experiments is therefore consistent with topological superconductivity in our samples. Similar ripples observed in a recent experiment on Bi$_2$Se$_3$/\Nbs~\cite{Bi2Se3_NbSe2_1} junctions may also share a similar origin.

While the appearance of super-gap ripples in the dI-dV spectrum strongly indicates the onset of topological superconductivity, the lack of a simple analytical expression to derive the positions of the ripples makes a direct comparison with experiments difficult. To enable such a comparison, band-structure calculations to identify the locations of the topological bands with respect to the trivial sub-bands will be required. Nonetheless, the mixing of the topological surface states with bulk-derived bands is {\it always} accompanied by the appearance of a 2D Majorana mode at zero energy making this a rather generic feature of topological superconductivity. Combined with its rather striking experimental signature, namely, the appearance of superconducting ripples at biases much larger than the superconducting gap voltage, this method will serve as a litmus test for topological superconductivity for future researchers.

\section{Conclusions}

In all, we have provided three distinct arguments in favour of an observation of topological superconductivity in our samples: i) The appearance of a two-gap spectrum, the larger gap associated with \Nbs~and the smaller arising from a proximity induced gap of the topological surface states ii) An intrinsic asymmetry in the conductance spectrum that arises due to the linear density of states of the topological surface state dispersion, and vanishes at temperatures and magnetic fields that destroy the proximity effect. iii) Differential conductance rippples at biases larger than the superconducting gap voltage, that are explained as a consequence of multi-band electron-hole mixing when topologically non-trivial superconducting order is induced. We propose that such features in differential conductance spectra can provide unambiguous tests for topological superconductivity. Our work therefore establishes the possibility of inducing superconductivity into the topological surface states of a bulk insulating topological insulator. This work demonstrates the possibility of realizing p-wave superconductivity at topological insulator/superconductor interfaces as originally envisaged by Fu and Kane~\cite{Fu2008}.

\section{Methods}

\BSTS single crystals are prepared using the modified Bridgman method. \Nbs single crystals are prepared using the chemical vapour transport method. Thin flakes of \BSTS and \Nbs are exfoliated using the scotch-tape method. Heterostructures of  NbSe$_2$/BSTS are prepared using the all-dry van der Waals transfer technique. E-beam lithography followed by e-beam evaporation of Cr/Au:10nm/100nm is used to define electrical contacts to the sample. Further details of device fabrication and single crystal characterization can be found in supplementary material section A and B. Electrical transport experiments are performed using a low-frequency AC lock in method. 10nA-20nA of AC bias currents are used throughout the experiments. For differential conductance measurements, the DC bias is generated using a 16-bit digital-to-analog converter, and is added to the AC bias within a home-built adder setup. All measurements reported here use a current biased scheme.

{\bf Acknowledgements}

The authors thank Diptiman Sen and Jainendra K. Jain for helpful and insightful discussions. A.B. thanks MHRD, Govt. of India for support. A.S. thanks KVPY, Govt. of India for support. P.S.A.K. thanks Nanomission, Department of Science and Technology, Govt. of India for support. The authors thank NNFC and MNCF, Centre for Nano Science and Engineering at the Indian Institute of Science Bangalore for fabrication and characterization facilities.

{\bf Supporting Information Available}

Supporting information contains details of sample fabrication and characterization, Blonder-Tinkham-Klapwijk fitting of differential conductance data, theoretical modeling of topological insulator/superconductor interfaces, and raw differential conductance data. This material is available online at \href{http://pubs.acs.org}{http://pubs.acs.org}.

\bibliography{SC_TI_acs_nano_v3.bbl}

\end{document}